\documentclass[twocolumn,superscriptaddress,reprint,nofootinbib,prd]{revtex4-1}%
\usepackage{amssymb}
\usepackage{color}
\usepackage{graphicx}
\usepackage{multirow}
\usepackage{dcolumn}
\usepackage{bm}
\usepackage[header,title,page,titletoc]{appendix}
\usepackage[utf8]{inputenc}
\usepackage{amsmath}
\usepackage{bbm}
\usepackage{amsfonts}%
\usepackage[bookmarks,colorlinks=false]{hyperref}
\newlength{\wordlength}

\newlength{\onewordlength}

    \newcommand{\ba}{\begin{eqnarray}}
    \newcommand{\ea}{\end{eqnarray}}
    \newcommand{\be}{\begin{equation}}
    \newcommand{\ee}{\end{equation}}

\newcommand {\bP} {{\mathbf P}}

\begin{document}

\title{Towards the understanding of $Z_c(3900)$ from lattice QCD}




\author{Chuan Liu}%
\email[Corresponding author. Email: ]{liuchuan@pku.edu.cn}
\affiliation{%
School of Physics and Center for High Energy Physics, Peking
University, Beijing 100871, China\\
Collaborative Innovation Center of Quantum Matter, Beijing 100871, China
}%

\author{Liuming Liu}
\affiliation{Institute of Modern Physics,Chinese Academy of Sciences, Lanzhou 730000, China\\
University of Chinese Academy of Sciences, Beijing 100049, China}






\author{Ke-Long~Zhang}
\affiliation{%
School of Physics, Peking University, Beijing 100871, China
}


%
 \begin{abstract}
 Within the framework of three-channel  Ross-Shaw effective range theory, we derive the
 constraints among different parameters of the theory in the case of a narrow resonance close
 to the threshold of the third channel,
 which is relevant for the resonance-like structure $Z_c(3900)$.
 The usage of these constraint relations,
 together with the multi-channel L\"uscher formula in lattice QCD calculations
 are also discussed and the strategies are outlined.
 \end{abstract}

 \maketitle



 \section{Introduction}
 \label{sec:intro}

 In the past decade, exotic hadronic resonance-like structures, known as $XYZ$ particles,
 have been discovered by various experiments, with $Z_c(3900)$ being a typical
 example~\cite{Ablikim:2013mio,Liu:2013dau,Xiao:2013iha}.
 The exotic structures have been discovered in both charm and bottom sectors
 which necessarily bear a four valence quark structure $\bar{Q}q\bar{q}'Q$ with $Q$ being a heavy-flavor
 quark while $q$ and $q'$ being two different light flavored ones.
 They also tend to appear close to the threshold of two heavy mesons
 with valence structure $\bar{Q}q$ and $\bar{q}'Q$.
 The physical nature of these structures have
 been contemplated and discussed in many phenomenological studies.
 For a recent review on these matters, see e.g. Ref.~\cite{Liu:2016ahj,Guo:2017jvc}.
 Despite many studies, the nature of $Z_c(3900)$ remains unclear.
 It is therefore highly desirable that non-perturbative studies
 like lattice QCD could provide some useful information.

 Contrary to the phenomenological studies, lattice studies on these states remain relatively scarce.
 A lattice study was performed by S.~Prelovsek et al. who investigated the energy levels
 of the two charmed meson system in the channel where $Z_c$ appearing
 in a finite volume~\cite{Prelovsek:2014swa}.
 They used quite a number of operators, including two-meson operators
 in the channel of $J/\psi\pi$, $D\bar{D}^*$ etc. and even tetraquark operators.
 However, no indication of extra new energy levels apart from the
 almost free scattering states of the two mesons.
 Taking $D\bar{D}^*$ as the main relevant channel, which is also supported by
 experimental facts, CLQCD utilized single-channel L\"uscher scattering formalism~\cite{luscher86:finitea,luscher86:finiteb,luscher90:finite,luscher91:finitea,luscher91:finiteb} to
 tackle the problem within single-channel approximation.
 They found slightly repulsive interaction between the
 two charmed mesons~\cite{Chen:2014afa,Chen:2015jwa}, making them unlikely to form bound states.
 A similar study using staggered quarks also finds no clue for the existence of the state~\cite{Lee:2014uta}.

 On the other hand,  HALQCD studied the problem using
 the so-called HALQCD approach~\cite{Ishii:2006ec} which is different from L\"uscher's adopted
 by the other groups mentioned above. An effective potential is first extracted from lattice
 data which is then substituted into the Schr\"odinger-like equation to solve for the scattering.
 They claimed that $Z_c(3900)$ can be reproduced and it is a structure formed due to strong cross-channel interactions among three channels, $J/\psi\pi$, $\eta_c\rho$ and $D\bar{D}^*$,
 see Ref.~\cite{Ikeda:2016zwx,Ikeda:2017mee} and references therein.
 This scenario will be referred to as the HALQCD scenario in the following.

 Recently, in order to clarify this mismatch of the two types of approaches,
 CLQCD performed a two-channel lattice study using the two-channel
 Ross-Shaw effective range expansion~\cite{Chen:2019iux}.
 They took the two channels $J/\psi\pi$ and $D\bar{D}^*$ that are most strongly coupled to $Z_c(3900)$.
 It is found that, in this two-channel approach, the parameters of the Ross-Shaw matrix
 do not seem to support the HALQCD scenario. The parameters turn out to be large and the
 Ross-Shaw $M$ matrix is far from singular, which is required for a resonance close
 to the threshold. However, since only two channels are studied, it is still not
 a direct comparison with the HALQCD approach in which three channels have been studied.
 In this paper, we move one step further to close this gap.
 We take exactly the same three channels as HALQCD did, namely $J/\psi\pi$, $\eta_c\rho$ and $D\bar{D}^*$.
 We utilize the Ross-Shaw effective range theory~\cite{Ross:1960len,Ross:1961ere} for the above mentioned three channels and derive the constraint relations among the parameters of Ross-Shaw matrix $M$
 (6 real parameters in total), assuming that there is a resonance close to the threshold of the third channel.
 Similar constraint relations in the two-channel case have been discussed in detail
 by Ross and Shaw long time ago, see e.g. Ref.~\cite{Ross:1961ere}.
 However, to our knowledge, the corresponding constraint relations in the three-channel case is still lacking,
 which will be established in this paper.
 These constraint relations can be further checked in future lattice simulations,
 the strategy of which will also be outlined in this paper.

 The difficulty with the multi-channel L\"uscher approach is two-fold
 which we briefly outline below:
 \begin{itemize}
 \item First, with the number of channels $n$ increasing, the number of unknown functions
 entering the $S$-matrix also increases rapidly. For example, in the case of two-channels,
 there are 3 functions while in the case of three channels, 6 functions
 are needed to describe the full $S$-matrix.
 On the other hand, L\"uscher formula only offers
 a single relation among these functions at a particular energy level,
 which is extracted from lattice simulation.
 Therefore, one needs to parameterize these unknown functions of energy
 in terms of constant parameters.
 Here, one could use the so-called $K$-matrix parameterization
 or the multi-channel effective range expansion advocated by Ross and Shaw,
 which is in fact a special case of the former. We will also take this choice
 in this paper.

 \item Second, the number of constant parameters
 needed to parameterize the $S$-matrix also grows quadratically
 fast when the number of channels $n$ is increased.
 Therefore, one should refrain to include too many channels.
 Based on the experimental facts and also hints from the HALQCD study,
 we focus on the three-channel L\"uscher approach in this paper.
 To be more specific, we will single out three most relevant channels for $Z_c(3900)$:
 $J/\psi\pi$, $\eta_c\rho$ and $D\bar{D}^*$, the first being
 the discovery channel for $Z_c(3900)$ and the second and the third have been shown to be
 dominant channels that couple to $Z_c(3900)$ by BESIII experiments~\cite{Ablikim:2015gda}.
 Similar to the single channel effective range expansion which
 is characterized by two real parameters, namely the scattering length $a_0$
 and the effective range $r_0$, in a three-channel situation,
 one needs 9 real parameters to describe the so-called Ross-Shaw matrix $M$:
 6 for the scattering length matrix and 3 for the effective range parameters.
 \end{itemize}

 This paper is organized as follows. In Section~\ref{sec:ross-shaw}, we briefly
 review the Ross-Shaw effective range expansion that is needed to parameterize $S$-matrix elements.
 In section~\ref{sec:resonance_in_RossShaw}, within the zero-range approximation
 of Ross-Shaw theory, we derive the constraint conditions that need to be satisfied in order
 to have a narrow resonance behavior close to the third threshold.
 These conditions are derived first in the limit where the coupling of the first
 two channels are switched off and then generalized to the case where it is turned on.
 In section~\ref{sec:strategy}, we briefly outline the strategies of the lattice computations
 and discuss how the constraints derived in Sec.~\ref{sec:resonance_in_RossShaw} can be tested.
 In Section~\ref{sec:conclude}, we will conclude with some general remarks.

 \section{The Ross-Shaw effective range theory}
 \label{sec:ross-shaw}

 In this section, we briefly recapitulate the Ross-Shaw effective range theory
 which is a generalization of usual effective range expansion to multi-channels.
 As already mentioned in the previous section, in order to utilize the multi-channel L\"uscher formula,
 it is  crucial to have a parameterization of the $S$-matrix elements in terms of constants
 instead of functions of the energy and the multi-channel effective range expansion
 developed by Ross and Shaw~\cite{Ross:1960len,Ross:1961ere} serves this purpose.

 In the single-channel case, this theory is just the well-know effective range expansion
 for low-energy elastic scattering,
 \be
 \label{eq:ere_single}
 k\cot\delta(k)=\frac{1}{a_0} +\frac{1}{2} r_0k^2+\cdots\;,
 \ee
 where $\cdots$ designates higher order  terms in $k^2$
 that vanish in the limit of $k^2\rightarrow 0$.
 Therefore, in low-energy elastic scattering, the scattering length $a_0$
 and the effective range $r_0$ completely characterize the scattering process.
 Ross-Shaw theory simply generalize the above theory to the case of multi-channels.
 For that purpose, they define a matrix $M$ via
 \be
 \label{eq:M_def}
 M=k^{1/2}\cdot K^{-1} \cdot k^{1/2}\;,
 \ee
 where $k$ and $K$ are both matrices in channel space.
 The matrix $k$ is the kinematic matrix which is a diagonal
 matrix given by
 \be
 \label{eq:kinematic_matrix_def}
 k=\left(\begin{array}{ccc}
 k_1 & 0 &0\\
 0 & k_2 &0\\
 0 & 0 & k_3\end{array}\right)\;,
 \ee
 and $k_1$, $k_2$ and $k_3$ are related to the scattering energy $E$.
 The matrix $K$ is called the $K$-matrix in scattering theory
 whose relation with the $S$-matrix is given by,
 \footnote{$K$-matrix is hermitian so that $S$-matrix is unitary.}
 \be
 \label{eq:SK_relation}
 S=\frac{1+iK}{1-iK}\;.
 \ee
 Another useful formal expression for the matrix $K$ is
 \be
 K=\tan\delta\;,
 \ee
 where both sides are interpreted as matrices in channel space.
 From the above expressions, it is easily seen that $K^{-1}$ that appears
 in Eq.~(\ref{eq:M_def}) is simply the matrix $\cot\delta$ and  without
 cross-channel coupling, the $M$-matrix is also diagonal
 with entries $M\sim\mbox{Diag}(k_1\cot\delta_1,k_2\cot\delta_2,k_3\cot\delta_3)$.
 Thus, it is indeed a generalization of the single channel case
 in Eq.~(\ref{eq:ere_single}).
 In their original paper, Ross and Shaw showed that the $M$-matrix
 as function of energy $E$ can be Taylor expanded around some
 reference energy $E_0$ as,
 \be
 \label{eq:ross-shaw-ere}
 M_{ij}(E)=M_{ij}(E_0)+\frac{1}{2}R_i\delta_{ij}\left[k^2_i(E)-k^2_i(E_0)\right]\;,
 \ee
 where we have explicitly written out the channel indices $i$ and $j$.
 The matrix $M_{ij}(E_0)\equiv M^{(0)}_{ij}$ is a real symmetric matrix in channel space that
 we will call the inverse scattering length matrix;
 $R\equiv\mbox{Diag}(R_1,R_2,R_3)$ is a diagonal matrix which we shall call
 the effective range matrix. $k^2_i$ are the entries for the kinematic
 matrix defined in Eq.~(\ref{eq:kinematic_matrix_def}).
 Therefore, for three channels, there are
 altogether 9 parameters to describe the scattering close to some
 energy $E_0$: 6 in the inverse scattering length matrix $M^{(0)}$
 and 3 in the effective range matrix $R$.
 One could further reduce the number of parameters to 6 by neglecting
 terms associated with effective ranges. This is called the zero-range
 approximation~\cite{Ross:1960len}.
 For convenience, we usually take $E_0$ to be the threshold of the
 third channel, i.e. that of $D\bar{D}^*$.

 It is understood that Ross-Shaw parameterization in
 Eq.~(\ref{eq:ross-shaw-ere}) is equivalent to the
 so-called $K$-matrix parameterization with three poles.
 In this $K$-matrix representation, assuming there are altogether $n$ open channels,
 the $n\times n$ $K$-matrix is parameterized as,
 \be
 \label{eq:K-matrix-poles}
 K(E)=k^{1/2}\cdot\left(\sum^n_{\alpha=1} \frac{\gamma_\alpha\otimes\gamma^T_\alpha}{E-E_\alpha}\right)
 \cdot k^{1/2}
 \;,
 \ee
 where $k$ is the kinematic matrix analogue of Eq.~(\ref{eq:kinematic_matrix_def}),
 the label $\alpha=1,2,\cdots,n$ designates the channels and each
 $\gamma_\alpha$ is a $1\times n$ real constant matrix (an $n$-component vector). It is shown
 in Ref.~\cite{Ross:1961ere} that this is equivalent to the effective range expansion~(\ref{eq:ross-shaw-ere})
 but the parameters are more flexible. In particular,
 $K$-matrix parameterization contains $(n^2+n)$ real parameters:
 $n^2$ for $n$ copies of $\gamma_\alpha$'s and another $n$ for the $E_\alpha$'s
 while an $n$-channel Ross-Shaw parameterization has $n(n+1)/2+n$ real
 parameters, $n(n-1)/2$ parameters less than the most general
 $K$-matrix given in Eq.~(\ref{eq:K-matrix-poles}). In this paper,
 we will focus on  the case of $n=3$ only.

\section{Resonance scenario in Ross-Shaw theory}
 \label{sec:resonance_in_RossShaw}

 In this section, we investigate the possibility of a narrow peak
 just close to the threshold of the third channel.
 In particular, this will be studied within
 the framework of three-channel Ross-Shaw theory.
 It turns out that this requirement will implement some
 constraints among the different parameters in Ross-Shaw theory.

 It is convenient to inspect the resonance scenario using the
 so-called $T$-matrix which is continuous across the threshold.
 Formally, it is related to the $K$-matrix via,
 \be
 K^{-1}=T^{-1}+i\;,
 \ee
 or equivalently as $T=K(1-iK)^{-1}$. The relation between the $S$-matrix
 and the $T$-matrix is given by,
 \be
 S=1+2iT\;,
 \ee
 where both $S$ and $T$ now are $3\times 3$ matrices in channel space.
 Since the scattering cross section $\sigma_{ij}$ is essentially proportional
 to $|T_{ij}|^2$, the so-called elastic cross section in a particular channel $i$
 is given by,
 \be
 \sigma_{ii}=\frac{4\pi}{k^2_i}|T_{ii}|^2\;.
 \ee
 Therefore, if we denote,
 \be
 w_{ii}\equiv \frac{T_{ii}}{k_i}=\frac{1}{\alpha_i(E)-i\beta_i(E)}\;,
 \ee
 with $\alpha_i$ and $\beta_i$ being real functions of the energy,
 then the elastic cross section in channel $i$ reads,
 \be
 \sigma_{ii}=4\pi |w_{ii}|^2==\frac{4\pi}{\alpha^2_i+\beta^2_i}\;.
 \ee
 Normally, the imaginary part of $w_{ii}$, namely $\beta_i(E)$, is a
 positive, smooth function of the energy in the energy region to be studied.
 In fact, if there were no coupling among different channels,
 we have $\beta_i=k_i$.
 The real part (i.e. $\alpha_i$), however, could develop a zero in
 the corresponding energy range, which then leads to a resonance peak structure.
 To be more specific, a resonance peak happens when $\alpha_i(E)=0$
 and the half-width positions for this peak can be obtained by
 the condition $\alpha_i(E)/\beta_i(E)=\pm 1$, respectively.

 To be more specific, the $T$-matrix in channel space looks like,
 \be
 \label{eq:T_general}
 T=k^{1/2}(M-ik)^{-1}k^{1/2}\;,
 \ee
 Therefore, if we define the matrix $w$ in channel space as,
 \be
 \label{eq:w_general}
 w=(M-ik)^{-1}\;,
 \ee
 the elements of which will be denoted by $w_{ij}$, then the
 following expression for $T_{11}$ can be obtained,
 \be
 \label{eq:w11}
 w_{11}\equiv \frac{T_{11}}{k_1}=\frac{1}{D}\left|
 \begin{array}{cc}
 M_{22}-ik_2 & M_{23}\\
 M_{23} & M_{33}-ik_3
 \end{array}\right|\;,
 \ee
 where $D$ is the determinant of the $3\times 3$ matrix,
 \be
 \label{eq:D_def}
 D=\left|\begin{array}{ccc}
 M_{11}-ik_1 & M_{12} & M_{13}\\
 M_{12} & M_{22}-ik_2 & M_{23}\\
 M_{13} & M_{23} & M_{33}-ik_3
 \end{array}\right|\;.
 \ee
 Similar expressions are obtained for $w_{22}$ and $w_{33}$.
 We get the following expression for $w_{ii}$ with $i=1,2,3$,
 \begin{widetext}
 \ba
 \label{eq:explicit_wii}
 w^{-1}_{11} &=& \alpha_1-i\beta_1=M_{11}-ik_1
 -M_{12}\frac{\left|\begin{array}{cc}
 M_{12} & M_{23}\\
 M_{13} & M_{33}-ik_3\end{array}\right|}
 {\left|\begin{array}{cc}
 M_{22}-ik_2 & M_{23}\\
 M_{23} & M_{33}-ik_3\end{array}\right|}
 +M_{13}\frac{\left|\begin{array}{cc}
 M_{12} & M_{22}-ik_2\\
 M_{13} & M_{23}\end{array}\right|}
 {\left|\begin{array}{cc}
 M_{22}-ik_2 & M_{23}\\
 M_{23} & M_{33}-ik_3\end{array}\right|} \;,
 \nonumber \\
  w^{-1}_{22} &=& \alpha_2-i\beta_2=M_{22}-ik_2
 -M_{12}\frac{\left|\begin{array}{cc}
 M_{12} & M_{13}\\
 M_{23} & M_{33}-ik_3\end{array}\right|}
 {\left|\begin{array}{cc}
 M_{11}-ik_1 & M_{13}\\
 M_{13} & M_{33}-ik_3\end{array}\right|}
 -M_{23}\frac{\left|\begin{array}{cc}
 M_{11}-ik_1 & M_{12}\\
 M_{13} & M_{23}\end{array}\right|}
 {\left|\begin{array}{cc}
 M_{11}-ik_1 & M_{13}\\
 M_{13} & M_{33}-ik_3\end{array}\right|}\;,
 \\
   w^{-1}_{33} &=& \alpha_3-i\beta_3=M_{33}-ik_3
 +M_{13}\frac{\left|\begin{array}{cc}
 M_{12} & M_{22}-ik_2\\
 M_{13} & M_{23}\end{array}\right|}
 {\left|\begin{array}{cc}
 M_{11}-ik_1 & M_{12}\\
 M_{12} & M_{22}-ik_2\end{array}\right|}
 -M_{23}\frac{\left|\begin{array}{cc}
 M_{11}-ik_1 & M_{12}\\
 M_{13} & M_{23}\end{array}\right|}
 {\left|\begin{array}{cc}
 M_{11}-ik_1 & M_{12}\\
 M_{12} & M_{22}-ik_2\end{array}\right|}\;.
 \nonumber
 \ea
 \end{widetext}
 In the above formulae, below a specific threshold, the corresponding momentum becomes
 pure imaginary. For example, below the threshold of the third channel, we have
 $-ik_3=\kappa_3$ with $\kappa_3$ being a positive real number.

 On the other hand, it is known from BESIII experiments~\cite{Ablikim:2013mio,Ablikim:2019ipd} that,
 close to the threshold of the third channel, all three elastic
 channels show resonant peaks. If we assume that these three peaks
 correspond to a single resonance structure, constraint equations can be obtained
 from Eq.~(\ref{eq:explicit_wii}). In the following, using Eq.~(\ref{eq:explicit_wii}),
 we will derive these equations that needs to be satisfied among the parameters.
 The corresponding conditions in the two-channel case has been studied long
 time ago by Ross and Shaw, e.g. Refs~\cite{Ross:1960len,Ross:1961ere}.
 However, to our knowledge, the case of three channels has not been studied explicitly
 which will be done within this paper.

\subsection{Resonance scenario in Ross-Shaw theory: $M_{12}=0$ case}
\label{subsec:m12zero}

 \P\ It is worthwhile to work in a somewhat simpler situation, namely
 that the coupling between channel $1$ and $2$ is negligible. This turns off
 the coupling between channel 1 and 2 completely, so we have $M_{12}=0$.
 Suppose that such a resonance structure arises from some pole structure
 in some Riemann sheet, then we could demand that the position of the pole
 to the be same, i.e. they corresponds to the same structure.

 In the limit where $M_{12}=0$, the condition $w^{-1}_{11}=w^{-1}_{22}=w^{-1}_{33}=0$
 turns out to yield a single equation (not three, but only one) for the parameters,
 \be
 \label{eq:complex_pole}
 M_{33}-ik_3=\frac{M^2_{13}}{M_{11}-ik_1}
 +\frac{M^2_{23}}{M_{22}-ik_2}\;,
 \ee
 where $k_1$, $k_2$ and $k_3$ are all related to the energy via,
 \be
 \label{eq:E_and_kis}
 \begin{aligned}
 E &=\sqrt{m^2_{J/\psi}+k^2_1}+\sqrt{m^2_{\pi}+k^2_1}\;,\\
   &=\sqrt{m^2_{\eta_c}+k^2_2}+\sqrt{m^2_{\rho}+k^2_2}\;,\\
   &=\sqrt{m^2_{D}+k^2_3}+\sqrt{m^2_{D^*}+k^2_3}\;.
   \end{aligned}
 \ee
 Here, $m_{J/\psi}$, $m_\pi$, etc. are the masses of the corresponding mesons
 and $k_i$'s with $i=1,2,3$ being the scattering momenta in various channels.
 Now, viewing the $k^2_i$'s, $i=1,2,3$ as complex variables that are
 related to each other by Eq.~(\ref{eq:E_and_kis}),
 one can solve Eq.~(\ref{eq:complex_pole}) in some Riemann sheet to yield
 the pole position for the  complex $k^2_i$'s. This pole then
 manifests itself as peaks in elastic cross sections in all three channels.
 Therefore, in the limit of $M_{12}=0$,
 the so-called HALQCD scenario is fully represented
 by Eq.~(\ref{eq:complex_pole}) in Ross-Shaw theory.

 To search for such solutions, we utilize the following notations.
 We assume that $k_3\equiv z$ is small in magnitude. Thus, we have,
 \be
 \delta E\equiv E-(m_{D^*}+m_D)=\frac{z^2}{2\mu_{DD^*}}\;,
 \ee
 with $\mu_{DD^*}$ being the reduced mass of $D$ and $D^*$.
 Similarly, $k_1$ and $k_2$ will assume their values at
 the third threshold, namely $k^{(0)}_1$ and $k^{(0)}_2$,
 plus small corrections that are linear in $z^2$.
 \be
 \begin{aligned}
 \delta k_1 &=\frac{z^2}{2\mu_{DD^*}(v^{(0)}_{J/\psi}+v^{(0)}_\pi)}=\gamma_1 z^2\;,
 \\
 \delta k_2 &=\frac{z^2}{2\mu_{DD^*}(v^{(0)}_{\eta_c}+v^{(0)}_\rho)}=\gamma_2 z^2\;,
 \end{aligned}
 \ee
 where $v^{(0)}_{J/\psi}$, $v^{(0)}_\pi$, $v^{(0)}_{\eta_c}$ and $v^{(0)}_\rho$
 are the speed of the corresponding mesons at the threshold. To be specific, we have,
 $v^{(0)}_{J/\psi}=k^{(0)}_1/E_{J/\psi}(k^{(0)}_1)$, etc.
 Therefore, the solution $z_0$, where all $w_{ii}$ diverge satisfy the following equation,
 \ba
 \label{eq:complex_pole2}
 M_{33}-iz_0 &=&\frac{M^2_{13}}{M_{11}-ik^{(0)}_1-i\gamma_1 z^2_0}
 \nonumber \\
 &+&\frac{M^2_{23}}{M_{22}-ik^{(0)}_2-i\gamma_2 z^2_0}\;.
 \ea
 This equation should be solved for small $|z_0|$ near the origin in
 the complex $z$ plane. Here smallness could be measured in some reasonable unit.
 A convenient choice is to use a unit system in which $k^{(0)}_1=1$
 adopted in Ref.~\cite{Chen:2019iux}.
 In such a system, every quantity in Eq.~(\ref{eq:complex_pole2}) becomes
 dimensionless and we are searching for $|z_0|\ll 1$ in the complex plane.

 Now, note that the l.h.s of Eq.~(\ref{eq:complex_pole2}) is linear in $z_0$
 while the r.h.s depends on $z^2_0$, therefore, we could write the solution $z_0$ as,
 \be
 z_0=z^{(1)}_0+z^{(2)}_0+\cdots\;,
 \ee
 where $z^{(i)}_0$ for different $i$ designates different orders of $z_0$, all of
 which are small, but the higher the index $i$ is, the even smaller the $z^{(i)}_0$ becomes.
 Taylor-expanding both sides of Eq.~(\ref{eq:complex_pole2}), order by order,
 we obtain the following equations,
 \begin{widetext}
 \ba
 iz^{(1)}_0 &=& \varepsilon\equiv M_{33}-\frac{M^2_{13}}{M_{11}-ik^{(0)}_1}-\frac{M^2_{23}}{M_{22}-ik^{(0)}_2}
 \label{eq:z_leading}\\
 z^{(2)}_0 &=& \left[\frac{M^2_{13}\gamma_1}{(M_{11}-ik^{(0)}_1)^2}
 +\frac{M^2_{23}\gamma_2}{(M_{22}-ik^{(0)}_2)^2}\right]\varepsilon^2
 \label{eq:z_next_leading}\\
 z^{(3)}_0 &=& 2i\left[\frac{M^2_{13}\gamma_1}{(M_{11}-ik^{(0)}_1)^2}
 +\frac{M^2_{23}\gamma_2}{(M_{22}-ik^{(0)}_2)^2}\right]^2\varepsilon^3
  \label{eq:z_next_next_leading}\\
 z^{(4)}_0 &=& \cdots \nonumber
 \ea
 \end{widetext}
 It is seen that the leading order equation~(\ref{eq:z_leading}) demands that
 the quantity $\varepsilon$ thus defined needs to
 be a complex number that is small in magnitude. Otherwise, there is no
 consistent small $z$ solution for Eq.~(\ref{eq:complex_pole2}). This implies that both the
 real part and the imaginary part has to be small.
 If we denote
 \be
 \varepsilon = \varepsilon_1 - i\varepsilon_2\;,
 \ee
 with both $\varepsilon_1$ and $\varepsilon_2$ being real,
 it is easy to work out the explicit expressions. It is also
 found that, the imaginary part parameter $\varepsilon_2>0$ at
 the threshold of the third channel. The sign of $\varepsilon_1$, however,
 is not definite, depending on other parameters.
 In order for them to be small, we have,
 \be
 \label{eq:closeness_condition}
 \begin{aligned}
 &\left| M_{33}-\frac{M^2_{13}M_{11}}{M^2_{11}+(k^{(0)}_1)^2}
 -\frac{M^2_{23}M_{22}}{M^2_{22}+(k^{(0)}_2)^2}\right| \ll 1\;,
 \\
 &\frac{M^2_{13}k^{(0)}_1}{M^2_{11}+(k^{(0)}_1)^2}
 +\frac{M^2_{23}k^{(0)}_2}{M^2_{22}+(k^{(0)}_2)^2} \ll 1\;.
 \end{aligned}
 \ee

 To leading order, the solution of the pole reads,
 \be
 z^{(1)}_0=-i\varepsilon=-\varepsilon_2-i\varepsilon_1\;,
 \ee
 which points out the approximate location of the pole position in the complex plane.
 To be more precise, the location is given by,
 \be
 \label{eq:z0}
 z_0=-\varepsilon_2-i\varepsilon_1+z^{(2)}+z^{(3)}+\cdots\;,
 \ee
 where $z^{(2)}$ and $z^{(3)}$ are given by Eq.~(\ref{eq:z_next_leading})
 and Eq.~(\ref{eq:z_next_next_leading}). More iterates can be obtained if necessary.

 We can now work out the elastic scattering cross sections close to the threshold
 of the third channel. These are given by Eqs.~(\ref{eq:explicit_wii}) by taking $M_{12}=0$.
 Taking e.g. the first channel, we have,
 \be
 w^{-1}_{11}=M_{11}-ik_1-\frac{M^2_{13}}{M_{33}-ik_3-\frac{M^2_{23}}{M_{22}-ik_2}}
 \;,
 \ee
 where $k_i$ takes real or pure imaginary values, depending on whether it is
 above or below the thresholds. Since the $k_i$'s are related to the total
 energy via Eq.~(\ref{eq:E_and_kis}), we know that the r.h.s vanishes when
 the $k_i$'s take complex values at $k_3=z_0$:
 \be
 M_{11}-ik_1(z_0)=\frac{M^2_{13}}{M_{33}-ik_3(z_0)-\frac{M^2_{23}}{M_{22}-ik_2(z_0)}}
 \;,
 \ee
 which is consistent with Eq.~(\ref{eq:complex_pole2}). Since the pole position
 is rather close to the third threshold, we expect that, large cross sections
 will be observed. Therefore, we introduce the function
 \be
 w^{-1}_{ii}=F_i(z)\;,
 \ee
 where in $F_i(z)$ the $k_i$'s are viewed as complex functions
 of $z$, which we still take as the complex $k_3=z$.
 We know that the function $F_i(z)$ has a zero at the location $z_0$ which
 is given by Eq.~(\ref{eq:z0}), and that $z_0$ is close to the origin.
 Therefore, we may expand,
 \be
 \begin{aligned}
 F_i(z) &=F_i(z_0)+F'_i(z_0)(z-z_0)+\cdots
 \\
 & \approx F'_i(0)(z-z_0)\;,
 \end{aligned}
 \ee
 where we have utilized the condition $F_i(z_0)=0$
 and $F'_i(z_0)\approx F'_i(0)$ since $z_0$ is rather close to the origin.
 Thus, the elastic cross section in channel $i$ reads,
 \be
 \label{eq:sigma_all}
 \sigma_{ii}=\frac{4\pi}{|F_i(z)|^2}=\frac{4\pi}{|F'_i(0)|^2|z-z_0|^2}
 \;,
 \ee
 which exhibits a typical resonance behavior. Here, it is understood
 that $z$ takes real or pure imaginary values, depending on whether
 it is above or below the third threshold. To be more explicit,
 if we take only the first approximation for $z_0$,  we have the following cross sections
 for above and below the third threshold,
 \be
 \label{eq:sigma_i_final}
 \sigma_{ii} =\left\{
 \begin{aligned}
 &\frac{4\pi}{|F'_i(0)|^2|[(k_3+\varepsilon_2)^2+\varepsilon^2_1]}\;,
 \\
 &\frac{4\pi}{|F'_i(0)|^2|[(\kappa_3+\varepsilon_1)^2+\varepsilon^2_2]}
 \;.
 \end{aligned}\right.
 \ee
 where the first/second line is for above/below the threshold,
 with $k_3=z=i\kappa_3$, $\kappa_3>0$ in the second case.
 Since we have $\varepsilon_2>0$, so the peak above the third threshold
 must be in the tail region. If $\varepsilon_1<0$, then we could see a
 full peak just below the threshold. If $\varepsilon_1>0$, however,
 a cusp will show up exactly at the threshold.

\subsection{Resonance scenario in Ross-Shaw theory: general case}
\label{subsec:m12nonzero}

 \P\ Here we would like to go beyond the approximation of $M_{12}=0$.
 We will show below that, the above results in fact hold in the most
 general case of three-channel scattering.

 For this purpose, we investigate Eq.~(\ref{eq:T_general}) and  Eq.~(\ref{eq:w_general})
 and realize that, in order to have a resonant behavior, the matrix $w=(M-ik)^{-1}$
 needs to be singular. This implies that the determinant $D$ defined
 in Eq.~(\ref{eq:D_def}) must vanish. Therefore, when viewed as a complex
 function of $k_3=z$, we may define,
  \be
 \label{eq:D_defz}
 D(z)=\left|\begin{array}{ccc}
 M_{11}-ik_1(z) & M_{12} & M_{13}\\
 M_{12} & M_{22}-ik_2(z) & M_{23}\\
 M_{13} & M_{23} & M_{33}-iz
 \end{array}\right|\;,
 \ee
 the complex resonance pole $z_0$ should be solved for under the condition of $D(z_0)=0$,
 in the neighborhood of the origin.
 In the above equation, functions $k_1(z)$ and $k_2(z)$ should be obtained
 by using the energy condition Eq.~(\ref{eq:E_and_kis}).
 Expanding both $k_1$ and $k_2$ around the origin we see that
 $k_{1,2}(z)=k^{(0)}_{1,2}+\gamma_{1,2}z^2$. Therefore, close to
 the origin, equation $D(z_0)=0$ yields a quintic equation for $z_0$.
 Since $|z_0|\ll 1$, we may expand the determinant $D(z_0)$ into a
 Taylor expansion. To first order, we get,
 \be
 D(z_0)\approx D(0)-(iz_0)\Delta_{33}(0)+\cdots\;,
 \ee
 where $\cdots$ designates terms higher orders in $z_0$ and
 $\Delta_{33}(0)$ being the cofactor for the matrix element $(M_{33}-ik_3)$
 in the $3\times 3$ matrix $(M-ik)$, i.e.,
 \be
 \Delta_{33}(0)=(M_{11}-ik^{(0)}_1)(M_{22}-ik^{(0)}_2)-M^2_{12}\;.
 \ee
 Therefore, to this order, the solution is
 \begin{widetext}
 \be
 \label{eq:z_leading_general}
 iz^{(1)}_0=\varepsilon\equiv \frac{D(0)}{\Delta_{33}(0)}=
 M_{33}+M_{13}\frac{\left|\begin{array}{cc}
 M_{12} & M_{22}-ik^{(0)}_2\\
 M_{13} & M_{23}\end{array}\right|}
 {\left|\begin{array}{cc}
 M_{11}-ik^{(0)}_1 & M_{12}\\
 M_{12} & M_{22}-ik^{(0)}_2\end{array}\right|}
 -M_{23}\frac{\left|\begin{array}{cc}
 M_{11}-ik^{(0)}_1 & M_{12}\\
 M_{13} & M_{23}\end{array}\right|}
 {\left|\begin{array}{cc}
 M_{11}-ik^{(0)}_1 & M_{12}\\
 M_{12} & M_{22}-ik^{(0)}_2\end{array}\right|}
 \ee
 \end{widetext}
 It is easy to verify that, in the limit of $M_{12}=0$, this reproduces
 the previous result in Eq.~(\ref{eq:z_leading}).
 The discussions about elastic scattering cross section remains unchanged.
 The only thing that needs to be modified is the explicit expression
 for the solution $z_0$ to various orders, which, to the first order,
 is now shown in Eq.~(\ref{eq:z_leading_general}) instead of Eq.~(\ref{eq:z_leading}).
 Again, higher order expressions can be obtained easily if necessary.

 \section{Multi-channel L\"uscher formula and the strategy for lattice computations}
 \label{sec:strategy}


 In this section, we briefly outline the strategies for a lattice calculation
 within the multi-channel L\"uscher approach for three channels.
 As we have mentioned in Sec.~\ref{sec:intro}, in the case of three-channels,
 one first needs a parameterization for the $S$-matrix in terms of functions, and
 furthermore in terms of the Ross-Shaw parameters.

 The most general form of $S$-matrix for three channels, assuming time reversal symmetry,
 was first given by Waldenstr{\o}m in 1974 and it looks like the following~\cite{Waldenstrom:1974zc},
 \be
 \label{eq:S2-canonical}
 S=\left[\begin{array}{ccc}
 \eta_1 e^{2i\delta_1} & iX_{12}e^{i(\delta_{12})} & iX_{13}e^{i(\delta_{13})}\\
 iX_{12}e^{i(\delta_{12})} & \eta_2 e^{2i\delta_2} & iX_{23}e^{i(\delta_{23})}\\
 iX_{13}e^{i(\delta_{13})} & iX_{23}e^{i(\delta_{23})} & \eta_3 e^{2i\delta_3} \\
 \end{array}\right]\;,
 \ee
 where $\delta_1$, $\delta_2$ and $\delta_3$ are scattering phases in channel 1, 2 and 3 respectively
 and $\eta_i\in[0,1], i=1,2,3$ are called the inelasticity parameters for each channel, all of
 which are functions of the energy.
 The other parameters $X_{ij}$ and $\delta_{12}$, $\delta_{23}$ and $\delta_{23}$ are related to the
 $\delta_i$ and  $\eta_i$ in a complicated manner hence also functions of the energy.
 Interested reader can consult Ref.~~\cite{Waldenstrom:1974zc} for details.
 These 6 functions of energy are then parameterized within Ross-Shaw theory
 in terms of 9 real parameters: 6 for the scattering length matrix $M$, 3 for
 the effective ranges. Note that $S$ matrix is related to the $T$ matrix via $S=1+2iT$ while
 the latter is further related to the Ross-Shaw $M$ matrix via Eq.~(\ref{eq:T_general}).

 The multi-channel L\"uscher formula has many forms. The most convenient one
  is the one that is directly related to the Ross-Shaw $M$-matrix,
 \be
  \label{eq:luscher_cube_three}
  \det\left[M-B^{(\bP)}\right]=0\;,
 \ee
 where the matrix $B^{(\bP)}$, called the box function by Colin Morningstar et al~\cite{Morningstar:2017spu},
 is a complicated but computable function involving modified zeta-functions that can be
 obtained from the energy eigenvalues in a finite box.
 The label $\bP$ designates the total three-momentum of the two-particle system
 so that it applies to also moving frames.
 The corresponding constraint equations that are derived in previous section
 needs to be boosted accordingly using an appropriate Lorentz transformation.
 The explicit expression for the box function reads:
 \begin{widetext}
 \be
 \langle J'm_{J'}L'S'a'|B^{(\bP)}|Jm_JLSa\rangle=-i\delta_{aa'}\delta_{SS'}
 (u_a)^{L+L'+1}W^{(\bP a)}_{L'm_{L'};Lm_L}(k^2_i)
 \langle J'm_{J'}|L'm_{L'},Sm_S\rangle
 \langle Lm_L,Sm_S|Jm_J\rangle\;.
 \ee
 \end{widetext}
 Here, $J$, $m_J$, $L$ and $S$ corresponds to total angular momentum quantum number,
 the third component of total angular momentum, orbital angular momentum and spin
 quantum number of the two-particle state. The index $a$ designates other quantum numbers, e.g. channel or
  isospin etc. The function $W^{(\bP a)}_{L'm_{L'};Lm_L}(k^2_i)$ involves zeta-functions and
 the arguments $k^2_i$ with $i=1,2,3$ represents the momenta in the corresponding channels
 which are related to the energy via Eq.~(\ref{eq:E_and_kis}).

 For a given set of parameters in Ross-Shaw matrix $M$, the
 multi-channel L\"uscher formula~(\ref{eq:luscher_cube_three})
 can be viewed as a equation for the energy eigenvalues that enters
 the equation via the box function $B^{(\bP)}$.
 Therefore, when solved numerically it yields a set of energy eigenvalues in the finite box.
 These energy levels can be compared with the real energy levels obtained from the lattice simulations.
 This comparison in turn yields an estimate for various $M_{ij}$'s in the Ross-Shaw matrix,
 as illustrated in Ref.~\cite{Chen:2019iux}.
 On the other hand, as we have obtained the conditions that need to be satisfied
 by these parameters in order to have a resonance peak close to the threshold of
 the third channel, c.f. Eq.~(\ref{eq:closeness_condition}),
 one can directly check if the lattice extracted parameters
 really support such a scenario or not, as was already done in the two-channel case
 in Ref.~\cite{Chen:2019iux}.

 It is interesting to note that, in the general Ross-Shaw theory,
 namely Eq.~(\ref{eq:ross-shaw-ere}) can be utilized to any energy region.
 In particular, if we investigate only the region close to the third threshold,
 it is good enough to use the zero range expansion.
 This sets all the effective ranges to zero, leaving us with only 6 parameters.
 In other words, if we focus on the energy region very close to the threshold,
 zero range approximation is always valid. Of course, by utilizing the multi-channel
 L\"uscher formula, other energy levels that are somewhat distant from the
 threshold enter the game (via fitting of $M_{ij}$'s), therefore there could
 be some deviations from the zero-range approximation. Still, extraction of
 the $M_{ij}$'s and check whether they satisfy the constraints as outlined
 in Eq.~(\ref{eq:closeness_condition}) offers a crucial test.
 This comparison will hopefully clarify, or at least shed some light on the differences from
 two different approaches so far: the HALQCD approach and the conventional L\"uscher approach.
 In fact, one could try to arrange a situation where as many as possible energy levels
 are close to the third threshold. In such a case, one could utilize the zero range approximation
 without any problem as long as one drops the energy levels that are too distant from the threshold.

 \section{Conclusions}
 \label{sec:conclude}

 To shed more light on the nature of the  resonance-like structure $Z_c(3900)$,
 lattice studies have been performed over the years. However, some puzzles still remain.
 The existing lattice studies fall into two categories: the ones using L\"uscher's approach
 and the ones using the HALQCD approach. The results from these two types of approaches
 are not consistent with each other as they should. This discrepancy needs to be clarified.

 In this paper,  we study the problem using the three-channel Ross-Shaw theory,
 which is the generalization of the effective range expansion. We have obtained
 the constraint conditions that needs to be satisfied by various parameters
 of the theory in order to have a narrow resonance close to the threshold of
 the third channel, a scenario that $Z_c(3900)$ realizes.
 We have pointed out that, combined with the multi-channel L\"uscher formula,
 a real lattice computation could be performed which will yield the results
 for these parameters and furthermore, one could check if these constraint relations
 are supported by the lattice results or not.
 We have also outlined the strategies of such a lattice simulations
 on how to extract these parameters in a more reliable fashion.
 Currently, we are working on the simulations details along the lines that are described here and
 we hope to report the results soon.

 \section*{Acknowledgments}

 This work is supported in part by the Ministry of Science and Technology of
 China (MSTC) under 973 project "Systematic studies on light hadron spectroscopy", No. 2015CB856702.
 It is also supported in part by the DFG and the NSFC through funds
 provided to the Sino-Germen CRC 110 ``Symmetries and the Emergence
 of Structure in QCD'', DFG grant no. TRR~110 and NSFC grant No. 11621131001.
 LL acknowledges the support from the Key Research Program of the Chinese Academy of Sciences, Grant NO. XDPB09.


%

 \end{document}